\documentclass[11pt,fleqn]{article}
\usepackage{amssymb}

\usepackage{amsmath}

\usepackage{amsmath,amssymb,amsthm}

\begin{document}

\begin{center}
{\Large \textbf{Infinite-dimensional symmetry for wave equation
with additional condition}}

\vskip 3pt {\large \textbf{Irina YEHORCHENKO~$^\dag$ and Alla
VOROBYOVA~$^\ddag$}}

\vskip 6pt {$^\dag$~Institute of Mathematics of NAS Ukraine, 3
Tereshchenkivs'ka Str., 01601 Kyiv-4, Ukraine} E-mail:
iyegorch@imath.kiev.ua

\vskip 3pt $^\ddag$~Mykolayiv Petro Mogyla State Humanities
University, Mykolaiv, Ukraine
\end{center}

    \begin{abstract} Symmetries for wave equation with additional conditions
    are found. Some conditions yield infinite-dimensional symmetry
    algebra for the nonlinear equation. Ansatzes and solutions
    corresponding to the new symmetries were constructed.
\end{abstract}

We discuss conditional symmetries of the Klein-Gordon equation
\begin{equation} \label{vorobyova:Fwave}
\Box u = F(x,u)
\end{equation}
for the real-valued function $u = u(x_0,x_1,x_2,\ldots,x_n)$, $x_0
=t$ is the time variable, $x_0$, $x_1$, $x_2$,\ldots,$x_n$ are
space variables, $n \neq 1$. $\Box u $ is the d'Alembert operator
\[
\Box u = - \frac{\partial^2 u}{\partial x^2_0} + \frac{\partial^2
u}{\partial x^2_1}+ \frac{\partial^2 u}{\partial x^2_2}
+\frac{\partial^2 u}{\partial x^2_3}.
\]

The general equation in the class (\ref{vorobyova:Fwave}) is not
invariant with respect to any operators, with only particular
cases having wide symmetry algebras \cite{vorobyova:FSerov-dA}.

The maximal invariance algebra of the equation
(\ref{vorobyova:Fwave}) with $F=F(u)$ (not depending on $x$) is
the Poincar\'e algebra $AP(1,n)$ with the basis operators
\begin{gather*}
 p_{\mu}  =  ig_{\mu \nu} \frac {\partial}{\partial
x_{\nu} }, \qquad J_{\mu \nu}  = x_{\mu} p_{\nu} -x_{\nu} p_{\mu},
\end{gather*}
where $\mu$, $\nu$ take the values $0$, $1$, $2$,\ldots, $n$; the
summation is implied over the repeated indices (if they are small
Greek letters) in the following way:
\begin{gather*}
x_{\nu } x_{\nu} = x_{\nu} x^{\nu} =  x^{\nu} x_{\nu} =x_0^2
-x_1^2 -x_2^2- \ldots - x_n^2,  \\ g_{\mu \nu }= {\rm
diag}\,(1,-1,-1,\ldots,-1).
\end{gather*}

The invariance algebras of the equation (\ref{vorobyova:Fwave})
will also include dilation operators for $F=\lambda u^k$ or
$F=\lambda \exp u$ and conformal operators for $F=\lambda
u^{\frac{n+3}{n-1}}$.

The maximal invariance algebra of the equation
(\ref{vorobyova:Fwave}) with $F=F(x^2,u)$ ($x^2=x_{\nu } x_{\nu}$)
is a subalgebra of the the Poincar\'e algebra $AP(1,3)$ whose
basis operators are Lorentz boosts $J_{\mu \nu}$.

Symmetry of the linear equation (\ref{vorobyova:Fwave}) with $F=0$
and $F=\lambda \exp u$ with $n=2$ is infinite-dimensional.

Similarity solutions for the equation (\ref{vorobyova:Fwave}) can
be found by symmetry reduction with respect to non-equivalent
subalgebras of its invariance algebras
\cite{{vorobyova:FSerov-dA},{vorobyova:Tajiri84},{vorobyova:FSS},
{vorobyova:FbarAsolDA90}}.

Here we present some examples of conditional invariance of the
equation (\ref{vorobyova:Fwave}) - the symmetry with an additional
condition being not $Q$-conditional symmetry.

The concept of conditional symmetry was derived and discussed in
the papers
\cite{vorobyova:OlverRosenau,vorobyova:FTsyfra,vorobyova:FZhdanovCond,
vorobyova:ClarksonKruskal,vorobyova:LeviWinternitz}, and later it
was developed by numerous authors into the theory and a number of
algorithms for studying symmetry properties of equations of
mathematical physics and for construction of their exact solutions
(see e.g.\cite{vorobyova:zhdanov&tsyfra&popovych99}). Here we will
work with the following definition of the conditional symmetry:

\vskip 3pt

\noindent {\bf Definition 1.} {\it The equation
$\Phi(x,u,\underset{1}{u},\ldots , \underset{l}{u})=0$, where
$\underset{k}{u}$ is the set of all $k$th-order partial
derivatives of the function $u=(u^1,u^2,\ldots ,u^m)$, is called
conditionally invariant \cite{vorobyova:FSS} under the ope\-rator
\[
Q=\xi ^i(x,u)\partial _{x_i}+\eta ^r(x,u)\partial _{u^r}\nonumber
\]
if there is an additional condition
\begin{equation}
G(x,u,\underset{1}{u},\ldots ,\underset{l_1}{u})=0,
\label{vorobyova:G=0}
\end{equation}
such that the system of two equations $\Phi=0$, $G=O$ is invariant
under the operator $Q$}.

If (\ref{vorobyova:G=0}) has the form $G=Qu$, then the equation
$\Phi=0$ is called $Q$-conditionally invariant under the operator
$Q$ \cite{vorobyova:FSS}.

These definitions of the conditional invariance of some equation
are based on what is in reality Lie symmetry (see e.g. the
classical texts
\cite{{vorobyova:Ovs-eng},{vorobyova:Olver1},{vorobyova:BlumanKumeiBook}})
of the same equation with a certain additional condition.
Conditional symmetries of wave equation are specifically discussed
in \cite{{YeVheat},{vorobyova:FSerov88},{vorobyova:BarMosk}}.

The equation (\ref{vorobyova:Fwave}) with an additional condition
\begin{equation}
x_\mu u_\mu + \alpha u = 0 \label{vorobyova:add1}
\end{equation}

with $\alpha \neq 0$ has the maximal symmetry algebra determined
by the operators
\begin{equation}
X=(-\frac{1}{\alpha} u^{\frac{1}{\alpha}}x_\mu \int{\Phi_u
u^{\frac{1}{\alpha}-1}}du + C_{\mu \nu} x_{\nu} +dx_\mu) p_{x_\mu}
+ \Phi \partial_u, \label{vorobyova:op1}
\end{equation}

with $\Phi=\Phi(u,u^{\frac{1}{\alpha}}x_\mu)$ being an arbitrary
function of its arguments.

$\partial_u$ designates the operator $\frac{\partial}{\partial
u}$.

With $\alpha = 0$ the corresponding algebra is generated by the
operator
\begin{equation}
X=x_0 \phi^\mu ( \frac{x_a}{x_0},u) p_{x_\mu} +
\psi(\frac{x_a}{x_0},u)
\partial_u, \label{vorobyova:op2}
\end{equation}

with $\phi^\mu, \psi$ being arbitrary functions of their
arguments.

The additional condition (\ref{vorobyova:add1}) can be presented
as $Du=0$, where $D$ is the dilation operator
\begin{equation}
D=x_\mu
\partial_\mu + i \alpha u \partial_u. \label{vorobyova:D}
\end{equation}

The equation (\ref{vorobyova:add1}) has the general solution
\begin{equation}
u=x_0^\alpha\phi( \frac{x_a}{x_0}). \label{vorobyova:anz1}
\end{equation}

where $\phi$ ia an arbitrary function. If we use
(\ref{vorobyova:anz1}) with $\omega_a=\frac{x_a}{x_0}$ as an
ansatz for
\begin{equation} \label{vorobyova:Lwave}
\Box u = 0,
\end{equation}

we get the reduced equation
\begin{equation}
(1+2\alpha)\omega_a\phi_{\omega_a}+\omega_a\omega_b\phi_{\omega_a\omega_b}
+\alpha(\alpha+1)\phi-\phi_{\omega_a\omega_a}=0.
\label{vorobyova:reduced1}
\end{equation}

Summation is implied over the repeated indices. The ansatz
(\ref{vorobyova:anz1}) corresponds to the operator
(\ref{vorobyova:D}) that is a Lie symmetry operator of the
equation (\ref{vorobyova:Lwave}).

We found some particular solutions of the equation
(\ref{vorobyova:reduced1}). If we put $\phi=\phi(\omega)$,
$\omega=m_a\omega_a$, $m_a$ are parameters with $m_am_a=1$, we get
an ordinary differential equation
\begin{equation}
(1+2\alpha)\omega\phi^\prime+(\omega^2-1)\phi^{\prime\prime}
+\alpha(\alpha+1)\phi=0. \label{vorobyova:reduced2}
\end{equation}

Its solution for $\alpha=0$ is
\[
\phi=c_1\ln|\omega+\sqrt{\omega^2-1}|+c_2, \nonumber
\]
for $\alpha=-1$ it is
\[
\phi=c_1(\frac{\omega}{2}\sqrt{\omega^2-1}-\frac{1}{2}\ln|\omega+\sqrt{\omega^2-1}|)+c_2.
\nonumber
\]

If $\phi=\phi(\omega)$, $\omega=\omega_a\omega_a$, $\alpha=0$,
then a solution of the equation (\ref{vorobyova:reduced1}) has the
form
\[
\phi=\int\omega^{-\frac{n}{2}}(\omega-1)^{-\frac{n}{2}-1}d\omega.
\nonumber
\]

The obtained solution are classical symmetry solutions of the
equation. However, by application of the group transformations
corresponding to the infinite-dimensional conditional symmetry
operators it is possible to multiply these solutions and to obtain
new ones that will not be classical symmetry solutions.

The conditional symmetries (\ref{vorobyova:op1}) and
(\ref{vorobyova:op2})  can also be considered as a hidden symmetry
of the equation (\ref{vorobyova:Lwave}), that is new symmetries of
the reduced equation (\ref{vorobyova:reduced1}) with $\alpha\neq
0$ or $\alpha=0$ that is not present for the original equation.

\vskip 3pt

\noindent {\bf Definition 2.} {\it An equation is said to have
hidden conditional invariance if a reduced equation is
conditionally invariant under some additional condition}
\cite{vorobyova:Yehorchenko2003}.

This definition stems from the definition of the hidden invariance
\cite{vorobyova:Abraham}.

Further we present an ansatz and solutions for the equation
(\ref{vorobyova:Lwave}) with another additional condition
\begin{equation}
x_\mu x_\nu u_{\mu \nu} + \alpha x_\mu u_\mu  = 0.
\label{vorobyova:add2}
\end{equation}

The condition (\ref{vorobyova:add2}) generates the following
ansatz for (\ref{vorobyova:Fwave}):
\begin{equation}
u=x_0^{1-\alpha}\psi( \frac{x_a}{x_0})+\phi(
\frac{x_a}{x_0})f(x_0), \label{vorobyova:anz2}
\end{equation}

where $f(x_0)=\ln x_0$ for $\alpha =1$ or $f(x_0)=1$ for $\alpha
\neq 1$.

This additional condition gives an ansatz leading to antireduction
\cite{vorobyova:antireduction}. There will be a system of two
reduced equations having the form:
\begin{gather}
2\omega_a\phi_{\omega_a}+\omega_a\omega_b\phi_{\omega_a\omega_b}-\phi_{\omega_a\omega_a}=0,
\nonumber \\
2\alpha\omega_a\psi_{\omega_a}+\omega_a\omega_b\psi_{\omega_a\omega_b}
-\psi_{\omega_a\omega_a}=0 \label{vorobyova:reduced3}
\end{gather}

for $\alpha \neq 1$, and
\begin{gather}
2\omega_a\phi_{\omega_a}+\omega_a\omega_b\phi_{\omega_a\omega_b}-\phi_{\omega_a\omega_a}=0,
\nonumber \\
\alpha\omega_a\psi_{\omega_a}+\omega_a\omega_b\psi_{\omega_a\omega_b}
-\psi_{\omega_a\omega_a}-\phi-2\omega_a\phi_{\omega_a}=0
\label{vorobyova:reduced4}
\end{gather}

for $\alpha =1$.

Let us adduce partial solutions of these reduced equations with
$\phi=\phi(\omega)$, $\omega=m_a\omega_a$, $m_a$ are parameters
with $m_am_a=1$

for (\ref{vorobyova:reduced3}) it is
\begin{gather*}
\phi=c_1 \ln \frac{\omega-1}{\omega+1}, \\
\psi=c_3\int \frac{d\omega}{(\omega^2-1)^\alpha},
\end{gather*}

for (\ref{vorobyova:reduced4}) it is
\begin{gather*}
\phi=c_1\ln{ \frac{\omega-1}{\omega+1}}, \\
\psi=\frac{1}{\sqrt{\omega^2-1}} \{
c_2\ln|\omega+\sqrt{\omega^2-1}|
-2\frac{c_1}{\sqrt{\omega^2-1}}+c_1\int\frac{1}{\sqrt{\omega^2-1}}
\ln|\omega+\sqrt{\omega^2-1}| d\omega\}.
\end{gather*}

Substituting the found solutions of the reduced equations into the
ansatz (\ref{vorobyova:anz2}), we obtain exact solution of the
equation (\ref{vorobyova:Lwave}).

\end{document}